\documentclass[prl,aps,preprint,amsmath,amssymb,showpacs,preprintnumbers]{revtex4-1}

\usepackage[pdftex]{color,graphicx}
\usepackage{bm}
\usepackage {ulem}
\usepackage{mathtools}
\usepackage[utf8]{inputenc}
\usepackage[T1]{fontenc}
\usepackage{lmodern}

\begin{document}

\title{Phase diagram of a three-dimensional
antiferromagnet with random magnetic
anisotropy}

\author{Felio A. Perez}
\altaffiliation{Current address: Integrated Microscopy Center, The University of Memphis, Memphis, TN 38152.}
\author{Pavel Borisov}
\author{Trent A. Johnson}
\author{Tudor D. Stanescu}
\author{David Lederman}
\email[]{david.lederman@mail.wvu.edu}
\affiliation{Department of Physics and Astronomy, West Virginia
University, Morgantown, WV 26506-6315, USA}
\author{M. R. Fitzsimmons}
\affiliation{Los Alamos National Laboratory, Los Alamos, NM 87545, USA}
\author{Adam A. Aczel}
\author{Tao Hong}
\affiliation{Quantum Condensed Matter Division, Neutron Sciences Directorate, Oak Ridge National Laboratory, Oak Ridge, TN 37831, USA}
\date{\today}

\begin{abstract}
Three-dimensional (3D) antiferromagnets with random magnetic anisotropy (RMA) experimentally studied to date do not have random single-ion anisotropies, but
rather have competing two-dimensional and three-dimensional exchange interactions which can
obscure the authentic effects of RMA. 
The magnetic phase diagram Fe$_{x}$Ni$_{1-x}$F$_{2}$ epitaxial 
thin films with true random single-ion anisotropy
was deduced from magnetometry
and neutron scattering measurements and analyzed using mean
field theory. Regions with uniaxial, oblique and easy plane
anisotropies were identified. A RMA-induced glass region was discovered where a
Griffiths-like breakdown of long-range spin order occurs.
\end{abstract}


\maketitle 
The behavior of insulating antiferromagnets (AFs) can range from spin
glass phases to random exchange, random
anisotropy, and random field
Ising models~\cite{fisher:1988}.  
As a consequence, these compounds
have received much attention due to their use as possible
experimental realizations~\cite{wong:1983PRB,vlak:1986} of
theoretical models~\cite{matsubara:1977JPSJ,fishman:1978PRB,matsubara:1979JPSJ} for
random magnets.  Although AFs with pseudo random magnetic anisotropy (RMA) have been studied
previously in Fe$_x$Co$_{1-x}$Cl$_2$, Fe$_x$Co$_{1-x}$Br$_2$, Fe$_x$Co$_{1-x}$TiO$_3$, and 
K$_{2}$Co$_{x}$Fe$_{1-x}$F$_{4}$ alloys, these systems only have approximate three-dimensional (3D)
order because the effective RMA actually consists of different intra- and inter-layer magnetic 
exchange coupling constants~\cite{vlak:1986,wilkinson:1959, tawaraya:1979, tawaraya:1980, wong:1980, wong:1983PRB, katsumata:1984,harris:1997}.  To see why this is important, consider the spin 
Hamiltonian 
\begin{equation}
H=\Sigma_i D\left(S_i^z\right)^2+\Sigma_{ij}\Delta J_{ij}S_i^zS_j^z+\Sigma_{ij}J_{ij}\mathbf{S}_i\cdot \mathbf{S}_j,
\label{eq:hgen}
\end{equation}
where $D$ is a single-ion anisotropy constant, $\Delta J_{ij}$ is the difference 
between intra- and inter-layer exchange coupling constants, and $J_{ij}$ is the intra-
layer exchange coupling constant.  In the mean field approximation, and taking into account only strongest neighbor interactions $J$, the Hamiltonian for a spin on the $\lambda$ sublattice of an antiferromagnet becomes
\begin{equation}
H_{\lambda}=D\left(S_\lambda^z\right)^2+z\Delta JS_\lambda^z\left<S_{\overline{\lambda}}^z\right>+zJ\mathbf{S}_\lambda\cdot \left<\mathbf{S}_{\overline{\lambda}}\right>,
\label{eqn:effA}
\end{equation}
where $z$ is the number of neighbors located on the sublattice $\overline{\lambda}$ that interact with a spin $S_\lambda$. The 
second term on the right hand side of Eq.~\ref{eqn:effA}, associated 
with an effective single-ion magnetic anisotropy resulting from the anisotropic 
exchange interaction, is strongly temperature dependent near the N\'eel temperature $T_N$ because
$\left<S^z_i\right>\rightarrow 0$ as $T\rightarrow T_N$.  On the other hand, the first term, 
which represents a true single-ion anisotropy,
is not temperature dependent and therefore dominates the physics in the vicinity
of the magnetic phase transition.  Consequently, the physics that governs a system with true random 
single-ion anisotropy near the phase transition will be in general different from 
the physics generated by an effective RMA produced by anisotropic exchange interactions.  
In this Letter, we report on the phase diagram of a solid solution of
two tetragonal 3D AFs that have orthogonal anisotropies originating solely from the single-ion anisotropies 
of each component, and thus represents a true 3D RMA antiferromagnet.

\begin{figure}
\includegraphics[width=8.5cm]{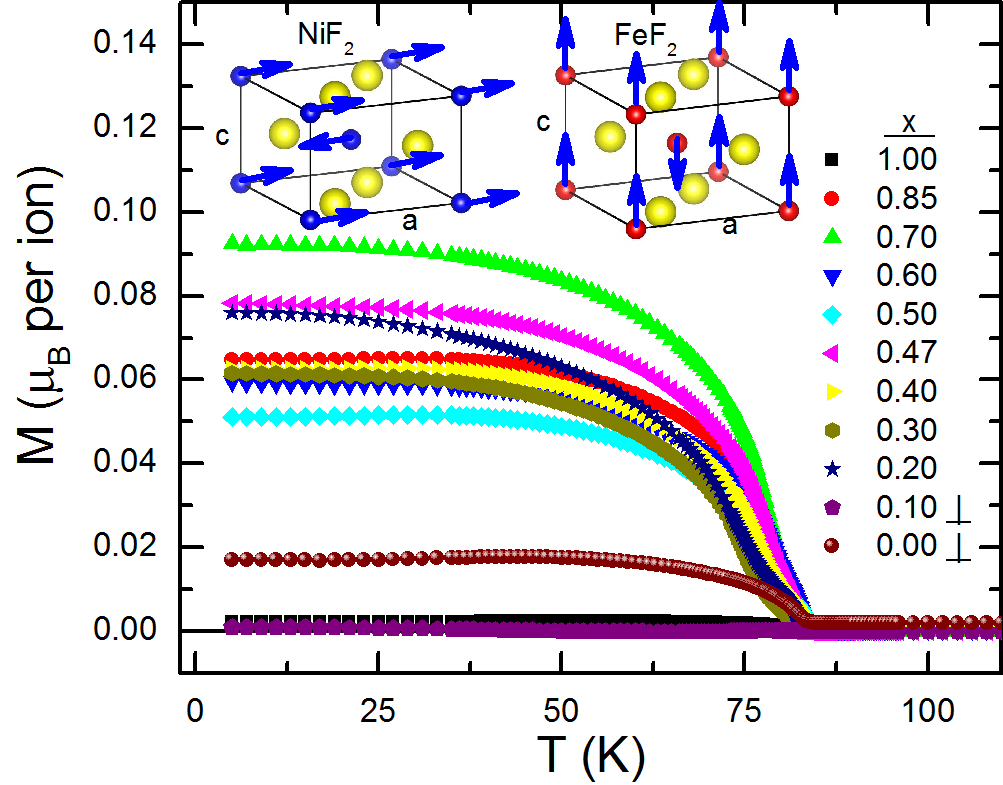}
\caption{\label{fig:Fig_M_vs_T} TRM for Fe$_x$Ni$_{1-x}$F$_2$ samples measured in $H=0$ after field
cooling in $H_{FC}=100$ Oe. Data for $x=0$ and 0.10 were measured with $H_{FC}$ perpendicular to the c-axis; all others measured with $H_{FC}$ parallel to the c-axis.  Inset: magnetic and crystalline structures of the
parent compounds NiF$_2$ and FeF$_2$. Yellow, blue, and red dots are F$^{+}$, Ni$^{2+}$, and red Fe$^{2+}$ ions, respectively.}
\end{figure}

FeF$_2$ and NiF$_2$ share the rutile 
crystal structure 
with similar lattice
parameters ($a=b=4.6974$ \AA\, $c=3.3082$ \AA\ for FeF$_2$ and
$a=b=4.6501$~\AA, $c=3.0835$~\AA\ for NiF$_2$ at room
temperature )~\cite{haefner:1966JAP,jauch:1993AC}.
Both materials are
3D AFs with similar exchange
interaction strengths $zJS(S+1)/3$ and thus they have similar
$T_N$s, 73.2K and
78.4K, for NiF$_2$ and FeF$_2$, respectively~\cite{hutchings:1970PRB, belanger:1987}. 
Their magnetic anisotropies are,
however, very different. FeF$_2$ has a strong
uniaxial anisotropy which results in its magnetic
moments being aligned along the tetragonal c-axis,  and is therefore considered an ideal realization
of the
3D Ising model~\cite{belanger:1987}. In NiF$_2$, moments
order antiferromagnetically in the a-b plane (Fig.~\ref{fig:Fig_M_vs_T}) and are canted by 
$\approx 0.4^\circ$ with respect
to the a- or b-axis~\cite{hutchings:1970PRB}. Weak ferromagnetism in NiF$_2$
is due to the presence of two non-equivalent magnetic sites in
the NiF$_2$ crystal lattice~\cite{moriya:1960PR}. 
The similarity of crystal structures and magnetic
exchange interactions in NiF$_2$ and FeF$_2$ suggests
that Fe$_{x}$Ni$_{1-x}$F$_{2}$ is an ideal system to
study RMA, which should vary from
transition metal site to site depending on whether it is
occupied by Ni$^{2+}$ (favoring a-b plane ordering) or Fe$^{2+}$
(favoring c-axis
ordering)~\cite{cheon:2007APL,fitzsimmons:2008PRB}.

\begin{figure}
\includegraphics[width=8.5cm]{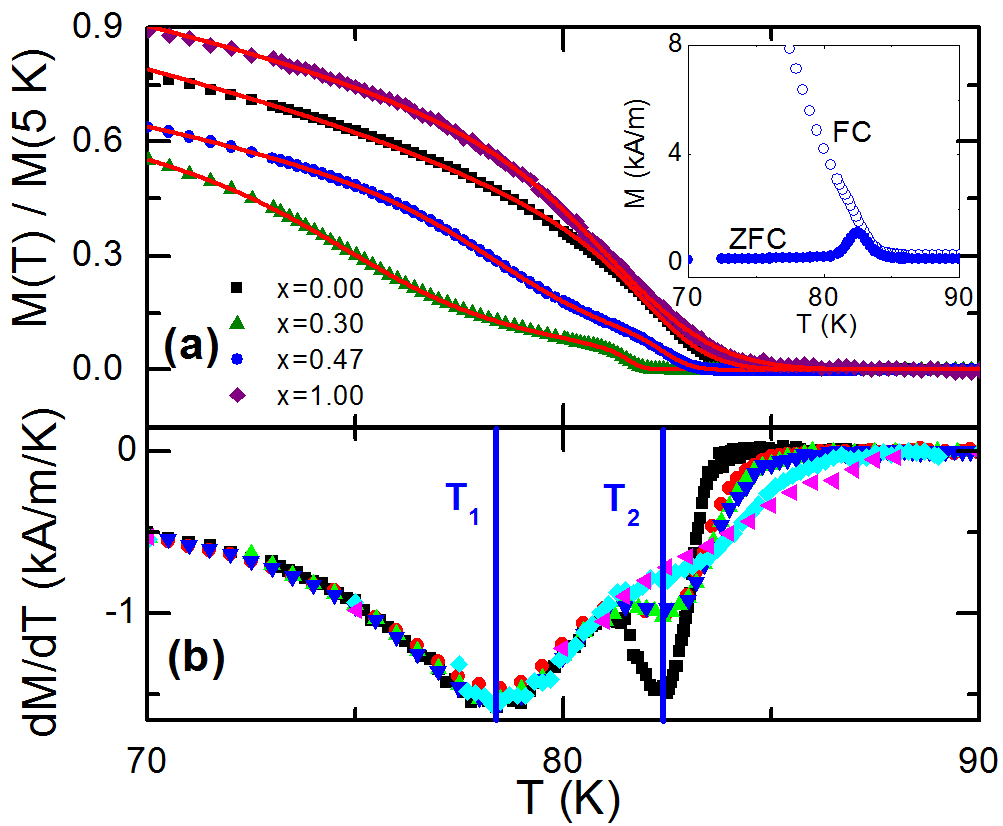}
\caption{\label{fig:dMdT} (a) TRM data ($H=0$) near the phase transition for four representative samples.  Symbols are data and red curves 
are fits to Eq.~\ref{eq:rounded} with two transitions for $x=0.30, 0.47$
and one transition for $x=1.0$ and $x=0.0$. 
Inset: $M$ measured while 
warming with $H=80$~Oe applied along the c-axis after ZFC
 from $T=300$~K to 5~K and during FC from 300~K to 5~K for the $x=0.47$ sample. 
(b) Numerical derivative $\partial M/\partial T$ of the $x=0.47$ TRM data 
 measured under different applied fields.  Vertical blue lines indicate transition temperatures $T_1$ and
$T_2$.}
\end{figure}

In order to study the 3D RMA anisotropy problem, epitaxial (110) Fe$_{x}$Ni$_{1-x}$F$_{2}$ films were grown with
nominal thicknesses of $37$ and $100$~nm on (110) MgF$_2$ 
substrates at $300$~$^{o}$C via molecular
beam epitaxy, as described
previously~\cite{cheon:2007APL,cheon:2007JAP}, and capped with a 10~nm BaF$_2$ 
or Pd layer to prevent oxidation.  The Fe concentration $x$ was determined 
using a quartz-crystal monitor 
with an accuracy of $\pm 0.05$~\cite{cheon:2007APL,cheon:2007JAP}. 
Thermal remanent
magnetization (TRM) measurements were carried out which consisted
measuring the magnetization $M$ while increasing $T$ from $T=5$~K after field-cooling (FC) from $T=300$~K in a field $H_{FC}=100$~Oe (Fig.~\ref{fig:Fig_M_vs_T}) along the in-plane $[001]$ (c-axis) and
 $[\bar{1}10]$ directions. The transition temperatures
were determined 
by fitting the data near the phase transition 
with a rounded power-law 
\begin{equation}
\label{eq:rounded}
I=\frac{I_0}{\sigma_c\sqrt{2\pi}}\int_0^\infty \left(1-T/T_c^\prime\right) ^{\beta}e^{-\left(T_c-T_c^\prime\right)^2/2\sigma_c^2}dT_c^\prime,
\end{equation}
where $T_c$ is a transition temperature, $\beta$ is a critical exponent, 
$\sigma_c$ is the width of the transition, and
 $I_0$ is an overall scaling factor~\cite{birgeneau:1998,lederman:1997PRB}.
  Magnetic hysteresis loops were measured as a function of $T$
and found to have large coercivities at low $T$ that decreased with increasing $T$ for $0.2<x<1$, in  agreement with previous measurements of Fe$_{x}$Ni$_{1-x}$F$_{2}$/Co bilayers (see Supplementary Materials)~\cite{cheon:2007APL,cheon:2007JAP}. FC and zero-field cooled (ZFC) measurements of $M$ vs. $T$ of all alloy samples behaved in a way that 
can be explained by the appearance of a ferromagnetic multi-domain
state during the ZFC process and its realignment after field-cooling (see Fig.~\ref{fig:dMdT} inset).
%


TRM data in Fig. \ref{fig:Fig_M_vs_T} show
the general effect of alloying on $M$. Relatively small deviations of $x$
 from the pure phases
result in significant increases of $M$ at low $T$, but these values
are much smaller than would be expected for ferrimagnetic
order~\cite{tsushima:1975}, and are therefore due to magnetic disorder. 

\begin{figure}
\includegraphics[width=8.5cm]{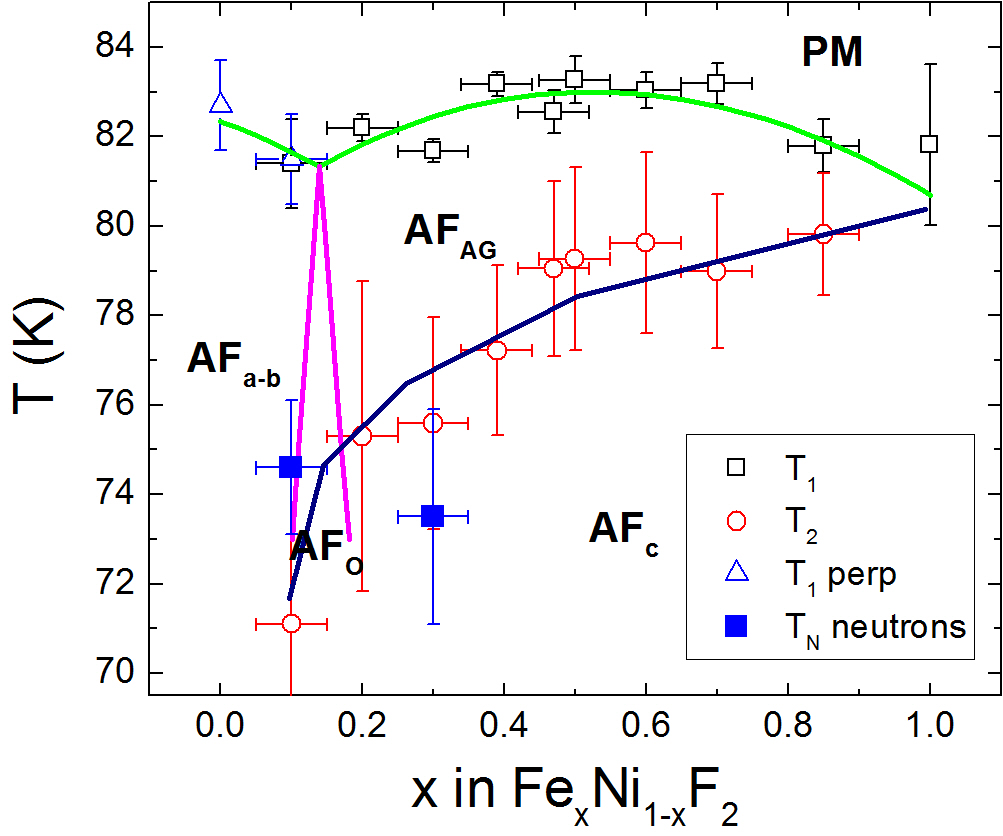}
\caption{\label{fig:PhaseD} Magnetic phase diagram for
Fe$_{x}$Ni$_{1-x}$F$_{2}$. Regions are indicated by AF$_{\text{a-b}}$ (ordering
in the a-b plane), AF$_\text{c}$ (ordering along the c-axis), AF$_\text{AG}$ (anisotropy glass phase), AF$_\text{O}$ (oblique phase), and PM (paramagnetic phase).  The PM-AF phase boundary calculated using MFT
 is denoted by the solid green curve.
$T_1$ and $T_2$ were determined from
fits to Eq.~\ref{eq:rounded} of TRM data. Horizontal error bars represent uncertainty in $x$ 
from quartz crystal monitor measurements.  Vertical error bars correspond to the transition widths 
$\sigma_{T_C}$.   Measurements taken for samples that had a response with $H\perp$
to the c-axis ($[\bar{1}10]$ direction, $\triangle$) are also indicated. Magenta lines enclose the MFT AF$_\text{O}$
region.  Transitions observed via neutron scattering are also indicated. The  dark blue curve is a $AF_\text{c}$-$AF_\text{AG}$ phase boundary drawn as a guide to the eye.}
\end{figure}

Examples of TRM phase transitions with  $H_{FC} ||$ c-axis, and $H_{FC}\perp$ c-axis for
the $x=0.0$ sample,
are shown in Fig.~\ref{fig:dMdT}(a).
%
The TRM data for all alloy samples
 had an inflection at a lower $T$ than
 the actual onset of the remanent magnetization, while the pure FeF$_2$ and NiF$_2$ samples 
only had one transition. The fits to the data using Eq.~\ref{eq:rounded} with two transitions for the alloys and one transition for the pure samples, indicated that $\beta\approx 0.34\pm0.05$ for all samples.  The transition temperatures and 
transition widths obtained from the fits for all samples are shown in Fig.~\ref{fig:PhaseD} and discussed further below.  The presence of two phase
transitions was more clearly seen in the form of two minima, at 
$T=T_1$ and $T=T_2$, in
the $\partial M/\partial T$ vs. $T$ data, as shown in Fig.
\ref{fig:dMdT}(b). When the TRM was measured in small $H$ applied along the c-axis,
the  transition at $T_2$ broadened substantially, while the 
transition at $T_1$ and the low temperature TRM remained unaffected.  
This phenomenon occurred for all samples with $0.2< x < 1.0$.
For $x=0.1$, a similar transition was observed with
$H\perp c$, indicating the existence of the easy-plane ordering similar
to that of pure NiF$_2$.  TRM data for $H || c$ had unusual behavior due to the
existence of an oblique phase, as discussed below (see Supplementary Materials).
 
The magnetic phase in the 
range $T_1<T<T_2$, where the magnetic structure is strongly coupled
to $H$, can be explained in two ways: (1) there is a first order spin-reorientation transition from
an Ising-like, single-axis anisotropy structure, similar to FeF$_2$, to a weakly ferromagnetic
structure, similar to NiF$_2$, at $T=T_1$ with increasing $T$, or (2)
the transition at $T=T_1$ is from the FeF$_2$ magnetic structure to a 
magnetically disordered structure.  In order to determine which of these explanations is correct, neutron scattering
was measured in $x=0.1$ and $x=0.3$ 100~nm thick samples 
using 3.0 and 3.4 meV neutron beams from the cold neutron triple-axis spectrometer 
(CTAX) at the High Flux Isotope Reactor, Oak Ridge National Laboratory (see Supplementary Materials for more details).  
Prior to measurement, the samples were cooled in $H_{FC}=60$~Oe $||$ c-axis.  
Once cooled to $T= 4$~K, $H$ was removed and the integrated intensities of the 
magnetic (100) and (001) reflections
with their background subtracted,  $I_{(100)}$ and $I_{(001)}$
(corresponding nuclear reflections are forbidden), were measured as a 
function of increasing $T$. 
From neutron scattering selection rules,  $I_{(100)}\propto L_\text{c}^2+L_\text{b}^2$, 
where $L_\text{b,c}$ is the component of the staggered magnetization vector $\mathbf{L}$ of the AF 
along the c- or b-axis, while $I_{(001)}\propto L_\text{ab}^2$, where 
$L_\text{ab}^2=L_\text{a}^2+L_\text{b}^2$ is the component of $\mathbf{L}$ in the $a-b$ 
plane.  The staggered magnetization 
vector is defined by $\mathbf{L}=\left(\mathbf{M_1}-\mathbf{M_2}\right)$, where 
$\mathbf{M_{1,2}}$ are the two 
sublattice magnetization vectors with $M_1=M_2$.   Explanation (1) would result in
  $I_{(001)}\neq 0$ only in the 
$T_1<T<T_2$ temperature range. On the other hand, explanation (2) requires
 that $I_{(100)},I_{(001)}>0$ only for $T<T_1$ because lack of long-range
order in the
$T_1<T<T_2$ range would preclude the observation of magnetic scattering.  

\begin{figure}
\includegraphics[width=8.5cm]{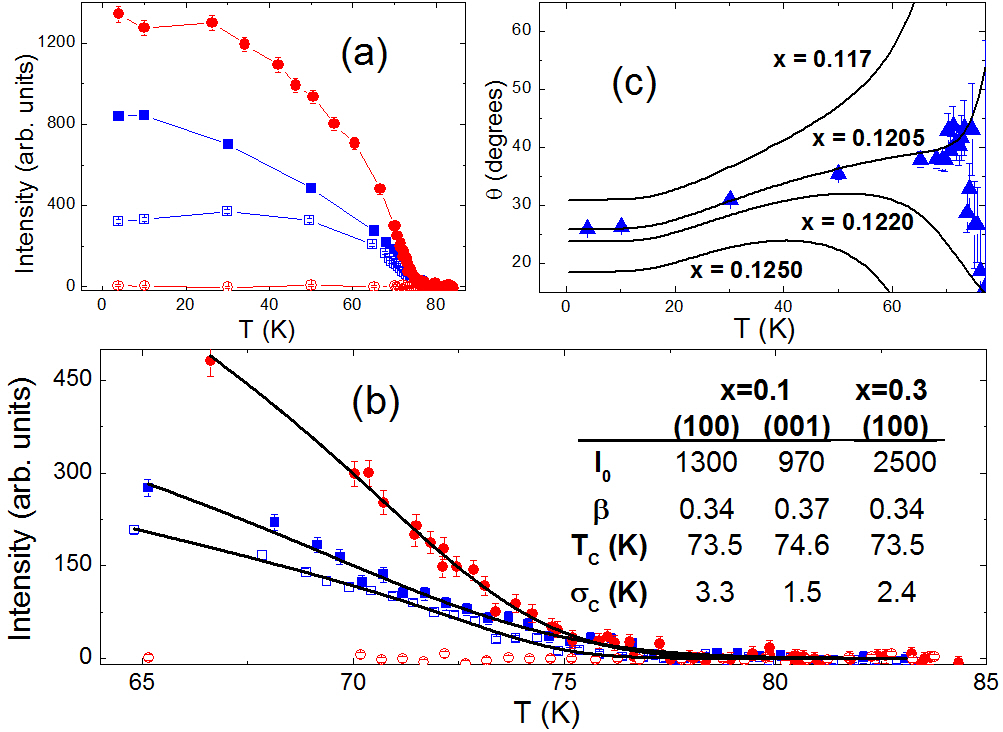}
\caption{\label{fig:neutrondata}
(a) Neutron scattering intensity as a function of temperature for the $x=0.3$ (red circles) and 
$x=0.1$ samples (blue squares).  Filled and open symbols indicate (100) and (001) reflections, respectively. Lines are guides to the eye. (b) Intensity near $T_c$.  Black curves are fits to  Eq.~\ref{eq:rounded} (with $\beta\rightarrow 2\beta$) and 
the resulting fitting parameters are shown in the graph.  (c) Angle of the staggered magnetization vector $\mathbf{L}$ as a function of$T$ with 
respect to the c-axis for the $x=0.1$ sample calculated from the data in (a).  The solid curves are
calculations using MFT for the values of $x$ indicated in the graph.}
\end{figure}

Figure~\ref{fig:neutrondata}(a) shows $I_{(100)} (T)$ and $I_{(001)} (T)$.
For both samples, the data indicate the presence of a single phase transition.
For the $x=0.3$ sample, $I_{(001)}=0$ for $0<T<85$~K, and therefore 
 the spins did not order antiferromagnetically in the a-b plane.  
For the $x=0.1$ sample, both $I_{(100)}$ and $I_{(001)}$ were non-zero at low 
$T$, and both $\rightarrow 0$ as  $T\rightarrow T_1$.  
This indicates
that $\mathbf{L}$ pointed in an oblique direction between the c-axis and the a-b plane.
Fitting the data to a rounded power law phase transition similar to
Eq.~\ref{eq:rounded}, but with $\beta\rightarrow 2\beta$ to take into account
the fact that $I\propto L^2$, yielded 
the results shown in Fig.~\ref{fig:neutrondata}(b).  The value of $T_N$ coincided with  $T_1$ measured for $x=0.1$ and $x=0.3$ samples within uncertainties.    
Because no significant intensity was observed for $T>T_N\approx 
T_1$ for either sample, we conclude that explanation (2) is correct:  there is a transition with increasing $T$ from 
an AF with long-range order to a disordered magnetic phase in the
$T_1<T<T_2$  range.  

The values of $\beta$ from neutron scattering agreed with those from the TRM measurements.  They are in better agreement with 
critical exponents corresponding the the 3D Ising, Heisenberg, and random exchange models ($\beta\approx 0.35$)~\cite{belanger:1987} than with the 3D random field  model ($\beta\sim 0.1$)~\cite{belanger:1991, birgeneau:1998,ye:2002}.  
To determine $\beta$ more accurately, 
and thus identify the transition's universality class, measurements must be made of the lineshape as a function of  scattering wavevector, $T$ and $H$ to 
take into account possible incoherent scattering backgrounds common in random magnetic systems~\cite{belanger:1991, birgeneau:1998}. Significantly thicker samples than the ones used here, possibly bulk 
single crystals, would be required. Accurately determining the universality class is therefore beyond the scope of this paper.

Whereas $L_{a-b}=0$ for the $x=0.3$ sample throughout the entire $T$ range, this is not the case for the 
$x=0.1$ sample.  This means that for $x=0.1$ an oblique phase exists throughout most, if not  the entire 
$T$ range, where $\mathbf{L}$ points at an angle $\theta$ away from the c-axis.
The value of $\theta$ can be determined using  
$L^2=L^2_{c}+L^2_{ab}$ and assuming that $L_b=L_a$, i.e., oblique domains are equally likely to
tilt towards the a- or b-axis, which yields
$\tan\theta = \left( I_{(001)}/I_{(100)}-1/2\right)^{1/2}$.
The dependence of $\theta$ on $T$ calculated from this equation 
is shown in Fig.~\ref{fig:neutrondata}(c).  




The phase diagram in Fig.~\ref{fig:PhaseD}, constructed
from the TRM and neutron scattering data, 
can be understood using mean field theory (MFT).
While MFT is inaccurate when predicting $T_N$, 
it is relatively successful at predicting quantities
which depend on changes in the effective field rather than on
its absolute value~\cite{wertheim:1969PR} and can describe, 
 at least qualitatively, the concentration dependence of
$T_N$ in mixed AF
systems~\cite{neda:1994PRB}. The 
spin Hamiltonian included
single-ion anisotropy terms and Heisenberg-type exchange
contributions, similar to the model used by
Moriya~\cite{moriya:1960PR} to study weak ferromagnetism in
NiF$_2$.  Using mean field decoupling for the exchange interactions while treating the single-site 
anisotropy terms exactly yields an average of the $\eta$ spin component for $\alpha$-type ions (either Fe or Ni) on the sublattice $\lambda$
\begin{equation}
\label{eq:spinavg}
\langle S^\eta_\alpha\rangle_\lambda = \frac{1}{Z_{\alpha\lambda}}\text{Tr}\left[ S^\eta_\alpha\exp\left(-H_{\alpha\lambda}/k_BT\right)\right],
\end{equation}
where the effective single-site Hamiltonian has the form
\begin{equation}
\label{eq:effH}
H_{\alpha\lambda}=\Sigma_\eta\overset{\approx}{h}^\eta_{\alpha\lambda}S^\eta_\alpha+D_\alpha\left(S^Z_\alpha\right)^2
\end{equation}
with the molecular field given by
\begin{equation}
\label{eq:mfeffH}
\overset{\approx}{h}^\eta_{\alpha\lambda}=z\Sigma_{\beta=\text{Ni,Fe}}J_{\alpha\beta}p_\beta\langle S^\eta_\beta\rangle_{\overline{\lambda}}.
\end{equation}
In Eq.~\ref{eq:spinavg} the partition function is $Z_{\alpha\lambda}=\text{Tr}\left[ \exp \left(-H_{\alpha\lambda}/k_BT\right)\right]$ and the spins $S^\eta_\alpha$ are represented by $3\times 3$ and $5\times 5$ matrices
for Ni and Fe ions, respectively.  
At a given $T$, $\langle S^\eta_\alpha\rangle_\lambda$  was determined numerically using an iterative scheme.  
Convergence was checked in the limit $D_\alpha =0$ by comparing with analytic expressions obtained 
within the full decoupling scheme (see Supplementary Materials for more details). 
The exchange coupling constants
$J_{\alpha_j\beta_k}$ were non-zero for next-nearest neighbor sites
$j$ and $k$ (between ions at the center of the tetragonal unit cell with those
at the corners)
and could take the values $J_\text{NiNi}$, $J_\text{FeFe}$, and
$J_\text{NiFe}$, corresponding to the different possible pairs
of spins.
Weaker exchange contributions with other neighbors 
 were neglected. The number of interacting neighbors was $z=8$ and $D_{\alpha}$ was positive for Ni ions and
negative for Fe ions.
Self-consistent calculations were carried out until the upper 
end of the phase diagram in Fig.~\ref{fig:PhaseD} was reproduced.  
 

Results of MFT calculations are shown in Fig.~\ref{fig:PhaseD}. 
The paramagnet (PM)-AF phase transition boundary was reproduced by adjusting the exchange constants to $J_\text{FeFe}=0.475$~meV,
$J_\text{NiNi}=1.63$~meV, and $J_\text{NiFe}=0.94$~meV, and using the known single-ion anisotropy constants $D_{Fe}=-0.80$~meV and $D_{Ni}=0.54$~meV \cite{hutchings:1970PRB,hutchings:1970JPC}.
$T_N$ values for pure NiF$_2$ and FeF$_2$ samples were larger than 
expected from the bulk parameters, but this has been previously attributed to 
strain (piezomagnetism)~\cite{mattson:1994JMMM,shi:2004PRB}.  The exchange constants were therefore different than 
the bulk values ($J_{\text{FeFe}}=0.451$~meV and $J_{NiNi}=1.72$~meV) \cite{hutchings:1970PRB,hutchings:1970JPC}. 
This non-monotonic dependence of $T_2$ on $x$ is due to an enhancement of the
exchange between unlike ions, 
$J_\text{FeNi}=0.88\text{ meV }>\sqrt{J_\text{FeFe}
J_\text{NiNi}}$, similar to what has been observed in Fe$_x$Mn$_{1-x}$F$_2$~\cite{wertheim:1969PR}.  
Increasing $J_\text{NiFe}$ much further shifts the minimum to
$x=0$.  

MFT also predicts a region 
where oblique ordering occurs, similar to prior MFT results for
 AF systems with anisotropic exchange couplings~\cite{matsubara:1977JPSJ,matsubara:1979JPSJ}. 
The canting angle $\theta (T)$ was calculated
using the same model and is depicted by the black curves in Fig.~\ref{fig:neutrondata}(c) (see SM).
The behavior was found to be extremely sensitive to $x$ and 
remarkably good agreement was found for $x\sim 0.1205$,  which is consistent with the
sample's nominal concentration of $x=0.1\pm 0.05$, but with $J_\text{NiFe}=1.02$~meV. This indicates
that other exchange interactions neglected by the model may play a role in determining $\theta (T)$.

Regions of different types of order predicted by the calculations are indicated
in Fig.~\ref{fig:PhaseD}. Whereas the calculated PM/AF boundary  
agrees well
with $T_2$,
neutron scattering data indicate that long-range order
disappears for $T>T_1$.  This leads to the conclusion that 
a
Griffiths-like~\cite{griffiths:1969,mccoy:1969,vojta:2006}
short-range order phase exists in the $T_1<T<T_2$ region 
as a result of the
random single-ion anisotropy.
Griffiths phases in other AFs
usually result from frustration of their
exchange interactions.
For example, magnetic field-induced antiferromagnetic correlations have been
reported in metamagnetic FeCl$_{2}$ \cite{binek:1994}, in
intraplanar frustrated FeBr$_{2}$\cite{binek:1995}, and in the dilute AFs
Fe$_{1-x}$Zn$_{x}$F$_{2}$~\cite{binek2:1995} and
Rb$_{2}$Co$_{1-x}$Mg$_{x}$F$_{4}$~\cite{binek:1998}. 
Here we propose a mechanism where a breakdown of magnetic long-range order
occurs at $T_{1}$,
with the random orthogonal single-ion magnetic anisotropy playing the
role of an effective local random field that leads to frustration. The emerging 
RMA-induced anisotropy glass region exists in the interval $T_{1}<T<T_{2}$, where $T_{2}$ is
the upper phase transition determined by the average exchange
interaction strength of the alloy.  The MFT used here is unable to reproduce this region
because it does not take into account local fluctuations of the effective field. 


In conclusion, the magnetic structure of Fe$_{x}$Ni$_{1-x}$F$_{2}$, an authentic 3D AF with random single-ion magnetic anisotropy, transforms from easy a-b
plane to the easy c-axis with increasing $x$
via an oblique phase region at $x=0.10-0.14$.
Two phase
transition temperatures, $T_1$ and $T_2$, were identified 
for $0.2<x<0.9$. 
Long-range order disappears for $T>T_1$, but 
short-range order
persists up to $T=T_2$.  The short-range order region is a result of the 
RMA which induces a magnetic
glass phase for $T_1<T<T_2$.  This phase is similar to magnetic glassy states 
formed as a result of combining structural disorder with frustrated exchange 
interactions, but with randomly distributed single-ion anisotropies replacing 
exchange frustration as the driving mechanism.  These effects have not been observed 
before in AFs because most AF systems studied to date do not have 
authentic single-ion 
RMA, but rather have an effective RMA induced by asymmetric exchange interactions 
which decreases rapidly as $T_N$ is approached.  

The supports of the National Science Foundation (grant
No.~DMR-0903861) and the WV Higher Education Policy Commission (Research Challenge Grant HEPC.dsr.12.29) at WVU
are gratefully acknowledged. Some of the work
was performed using the
WVU Shared Research Facilities.   Research conducted at ORNL and LANL was sponsored by the Scientific User Facilities Division, Office of Basic Energy Sciences, US Department of Energy.

%

\end{document}